\documentclass[preprintnumbers,amsmath,amssymb]{revtex4}
\usepackage{epsfig}
\usepackage{bm}

\newcommand{\s}{\sigma}
\newcommand{\la}{\lambda}

\begin{document}

\title{Classical evolution of fractal measures on the lattice}

\author{N.~G.~Antoniou}
\author{F.~K.~Diakonos}
\author{E.~N.~Saridakis}
\email{msaridak@phys.uoa.gr}
\author{G.~A.~Tsolias}
\affiliation{Department of Physics, University of Athens, GR-15771
Athens, Greece}

\date{\today}

\begin{abstract}

We consider the classical evolution of a lattice of non-linear
coupled oscillators for a special case of initial conditions
resembling the equilibrium state of a macroscopic thermal system
at the critical point. The displacements of the oscillators define
initially a fractal measure on the lattice associated with the
scaling properties of the order parameter fluctuations in the
corresponding critical system. Assuming a sudden symmetry breaking
(quench), leading to a change in the equilibrium position of each
oscillator, we investigate in some detail the deformation of the
initial fractal geometry as time evolves. In particular we show
that traces of the critical fractal measure can sustain for large
times and we extract the properties of the chain which determine
the associated time-scales. Our analysis applies generally to
critical systems for which, after a slow developing phase where
equilibrium conditions are justified, a rapid evolution, induced
by a sudden symmetry breaking, emerges in time scales much shorter
than the corresponding relaxation or observation time. In
particular, it can be used in the fireball evolution in a
heavy-ion collision experiment, where the QCD critical point
emerges, or in the study of evolving fractals of astrophysical and
cosmological scales, and may lead to determination of the initial
critical properties of the Universe through observations in the
symmetry broken phase.

\end{abstract}

\maketitle

\section{Introduction}

Chains of non-linear coupled oscillators are of fundamental
nature: they provide a laboratory to explore the setup of
thermodynamical properties through the microscopic dynamics in
complex systems \cite{Toda89,Pettini,Parisi97}. In addition, being
the discrete version of field theories, they naturally emerge in
any numerical study of the non-linear dynamics as well as
statistical mechanics of classical fields \cite{Wetter99}. One of
the most important questions in the later case is to determine the
conditions which can drive the evolving system towards a
thermalized stationary state. In the early days Fermi, Pasta and
Ulam \cite{Fermi55} have obtained deviations, even for large
times, from the naively expected equipartition of the energy among
the different oscillators. Through the efforts to explain these
results, it became clear that, for appropriate initial conditions,
a variety of stable periodic solutions (breathers, solitary waves)
\cite{Ford92}, defined on the non-linear chain, exists. Therefore,
the choice of the ensemble of the initial configurations strongly
influences the long time behavior of the system dynamics. Recent
works \cite{Parisi97,Wetter99} show that for a random ensemble of
initial configurations a sufficiently large system relaxes to the
usual equilibrium distribution, but the corresponding relaxation
time strongly depends on the parameters of the theory. These
studies include the case when a chain of oscillators is replaced
by a multi-dimensional ($2-D$ or $3-D$) lattice \cite{Wetter99}.
In fact, when considering non-linear lattices in more than one
dimensions, the thermalized equilibrium state can posses critical
properties. The question of how critical properties can
dynamically occur in a system of coupled oscillators is not yet
fully understood. Some recent investigations \cite{Pettini}
indicate that changes in the topological properties of the phase
space of the considered system are induced by fine tuning the mean
kinetic energy of the oscillators. The critical state, when
formed, is associated with appropriately defined fractal measures
on the non-linear lattice.

In the present work we investigate the scenario when such a
critical state has already been formed and a sudden symmetry
breaking drives the system apart from critical behavior.
Furthermore, it is assumed that the time scale of the symmetry
breaking process is much smaller than the relaxation time of the
oscillator dynamics. Although we are concerned in the general case
of the evolution of a set with fractal geometry, we focus in the
large subclass where the corresponding measure is generated by a
scalar field. In fact, the scalar field dynamics considered in
this work, covers a large variety of phenomena with critical
fluctuations which at equilibrium are described by the
universality class of the $3-D$ Ising model and its projections to
lower dimensions. Hence, the above model is of great physical
interest since it is realized in many different areas of physics.
In a heavy-ion collision experiment, if the fireball passes near
the QCD critical point it acquires critical correlations. The
subsequent expansion and cooling induces a non-equilibrium
evolution for the effective field equations of motion, which
dilutes the initial fractality. The question is weather the
corresponding relaxation time is long enough in order to acquire
imprints of the initial critical state at the detectors
\cite{manosqcd}. On the other hand, within the framework of scalar
field dynamics, one can impose initial conditions associated with
self-similar fluctuations of the inflationary field, and study
their evolution as they grow and transform to the large-scale
inhomogeneities of the observable Universe
\cite{inflation.fractal}. A similar approach, but with different
dynamics, has been applied to the fractal-like
structure of the Universe at large scales (star and galaxy
clusters). The evolution in this case is crucial for the
substructure survival or deformation
\cite{star.fractal,galaxy.fractal}.

In order to implement the aforementioned scenario we will use the
critical state as initial condition posed on the oscillators in
the lattice. The corresponding fractal measure is generated
through a suitable excitation of the oscillators. Contrary to the
existing analysis of correlations and their evolution in a fractal
lattice \cite{Marini97}, our approach is closer to the conditions
expected to occur in a real critical system where inhomogeneities
in the order parameter density have a fractal profile embedded in
a conventional space. The evolution of initial fractal measures
has been studied both classically, in the context of
reaction-diffusion models \cite{Alemany97}, as well as
quantum-mechanically \cite{Wojcik2000}. In the classical case an
unconventional decay of correlations was observed, while in the
quantum one the initial fractal dimension turns out to be a
conserved quantity.

We investigate first the $1-D$ case of a nonlinear chain and then
extend our treatment to lattices in higher dimensions. The $1-D$
example, although it cannot be directly related to a critical
system in the absence of long ranged interactions, due to the
no-go theorem of Peierls \cite{Peierls}, it helps for the simpler
illustration of the basic dynamical mechanisms which dominate the
evolution of the system in the meta-critical phase. The extension
to $2-D$ is straightforward using the insight gained by the $1-D$
model. The main finding of this work is a set of conditions which
control both qualitatively and quantitatively the time-scale for
which traces of the initial critical state, characterized by the
fractal mass dimension, sustain. In addition, we show that this is
a new time-scale not directly associated with the relaxation time
towards the false vacuum. Our analysis shows that the fractal
dimension, describing the geometry of the critical state, is a
valuable observable which can be determined even in the symmetry
broken phase, allowing for the calculation of critical exponents
and consequently for the determination of the universality class
of the occurring transition \cite{stinchcombe}.

The paper is organized as follows: in section II we describe the
dynamical model used in our analysis and we explain the algorithm
used to generate the initial conditions. In section III we give
the numerical solution of the equations of motion for the $1-D$
case as well as the observables which are relevant in quantifying
the effect of the sudden symmetry breaking. To facilitate our
analysis we first consider in subsection IIIa the harmonic chain,
and in subsection IIIb we include non-linear chain interactions.
In section IV we extend our study to $2-D$, discussing also the
higher dimensional case. Finally in section V we present briefly
our concluding remarks.

\section{Generating a fractal measure in a chain of non-linear oscillators}

The considered dynamical system consists of a set of coupled
oscillators located on an equidistant lattice. These oscillators
are the discretized version of a self-interacting scalar field. In
a simplified approach we investigate first the $1-D$ case when the
oscillators are arranged in a closed chain. The Lagrangian of this
system is taken as:
\begin{equation}
L_=\sum_{i=1}^{N}\,\left\{\frac{1}{2}\dot{\s}_i^2-
\frac{1}{4\alpha^2}\left[\left(\s_{i+1}-\s_{i}\right)^2+
\left(\s_{i}-\s_{i-1}\right)^2 \right]-V(\s_i)\right\},
\label{eq:leap-froglagr}
\end{equation}
where $\alpha$ is the lattice spacing and $V(\sigma_i)$ is the
self-interaction term. The coupling term between the oscillators
originates from the discretization of the spatial derivative term
in the lagrangian density of a scalar field $\s(x,t)$, while the
choice of the potential is adapted to the typical form for the
description of spontaneous symmetry breaking in second order phase
transitions.

 We study the dynamics implied by eq.
(\ref{eq:leap-froglagr}) using appropriate initial conditions
$\{\s_i(0)\}$ in order to define a fractal measure on the $1-D$
lattice \footnote{In fact an ideal fractal in the mathematical
sense cannot be defined on a discrete space. However, physical
fractals are always defined between two scales and therefore can
be embedded in a lattice.}. The construction of these
configurations is based on a finite approximation to the $1-D$
Cantor dust with prescribed Hausdorf fractal dimension $D_f$
\cite{Mandel83,Vicsek}. Although the construction algorithm for
such a set is described in several textbooks on fractal geometry
we also present it here briefly in order to be self-contained. The
first step in the algorithm is the partition of the finite real
interval $[0,1]$ into three successive subintervals of sizes
$\ell_1$, $\ell_2$ and $\ell_3$, in ratios
$\frac{\ell_1}{\ell_2}=\frac{r}{1-2r}$ and
$\frac{\ell_2}{\ell_3}=\frac{1-2r}{r}$, where
$r\in(0,\frac{1}{2})$, covering the entire set $[0,1]$. The middle
subinterval is called { \it trema} where the two others are called
{\it precurds} \cite{Mandel83}. At each step of the algorithm the
trema is omitted, while the precurds are divided also in three
subintervals with sizes fulfilling the same ratios as above. At
the $k$-th stage of the algorithm there are $2^k$ precurds and
$2^k-1$ tremas. In the limit $k\rightarrow\infty$ we obtain a
Cantor dust with fractal dimension $\log2/\log(1/r)$. Thus
choosing $r=2^{-1/D_f}$, we can practically construct a set with
the desired fractal dimension $D_f$.

A finite approximation to the Cantor dust with dimension $D_f$ can
be obtained using the centers of the precurds $x^{(k)}_i$
($i=1,..,2^k$) at the $k$-th algorithm step. This set can be
easily embedded on a finite equidistant lattice using the
transformation $\nu_i^{(k)}=\left[\frac{d_{max}}{d_{min}}\,
x^{(k)}_i\right]+1$, where $d_{min}=\min_{i\neq j}\vert
x^{(k)}_i-x^{(k)}_j\vert$ and $d_{max}=\max\vert
x^{(k)}_i-x^{(k)}_j\vert$, and $[..]$ denotes the integer part.
Thus, the set of $\nu_i^{(k)}$ is a realization of the Cantor dust
defined in the interval $(0,N\alpha]$, with
$N=\left[\frac{d_{max}}{d_{min}}\right]+1$. In this interval we
can construct a density as:
\begin{equation}
\rho_C^{(k)}(y)=\frac{1}{2^k} \sum_{i=1}^{2^k} \delta
(y-\nu^{(k)}_i). \label{eq:cantden}
\end{equation}
The fractal properties of the set are quantitatively depicted in
the scaling law:
\begin{equation}
M(\varepsilon)\sim\varepsilon^{D_f}, \label{eq:scaleprop}
\end{equation}
where $M(\varepsilon)$ is the number of set points $\nu_i^{(k)}$
within a distance $\varepsilon$ from any given reference point
$\nu_j^{(k)}$ ($j\neq i$) belonging to the set.

Using the density (\ref{eq:cantden}) we can map the fractal
geometry of the Cantor dust approximation to the non-linear
oscillators on the lattice by assuming that the displacement of
the $\nu$-th oscillator is obtained through the integral
\begin{equation}
\s_\nu=\eta_\nu2^k\int_{\nu-\delta/2}^{\nu+\delta/2}\,\rho_C^{(k)}(\xi)
d\xi, \label{eq:svalues}
\end{equation}
where $\eta_\nu$ is a random variable taking the values $\pm1$
with equal probability, $0<\delta\ll1$ and $\nu=1,..,N$. With this
choice it is straightforward to define a fractal measure on the
oscillator chain through the obviously fulfilled property
\begin{equation}
m(\zeta)=\langle\sum^{\nu=\nu_i^{(k)}+\zeta}_{\nu=\nu_i^{(k)}}
|\s_\nu|\rangle\propto\zeta^{D_f}, \label{fracmassdim}
\end{equation}
where the average is taken over all $\nu_i^{(k)}$.

\begin{figure}[h]
\begin{center}
\mbox{\epsfig{figure=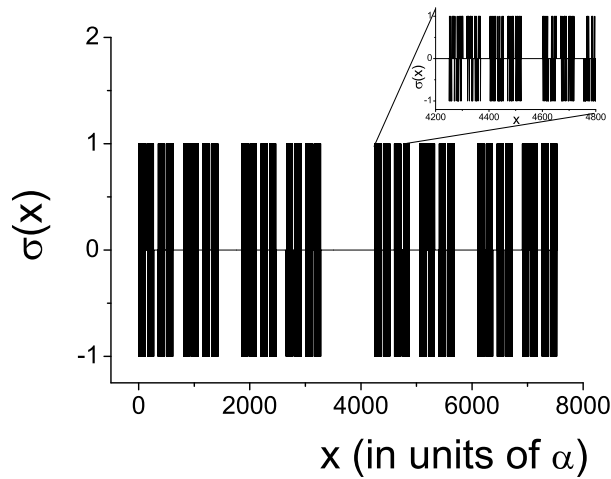 ,width=9cm,angle=0}}
\caption{\it The $\s$-field for a Cantor-like lattice of size
$N=18819$, obtained using $k=11$ and $r=2^{-6/5}$.}
\label{Det.sav}
 \end{center}
 \end{figure}

An example of a $\s$-field configuration is depicted in
fig.~\ref{Det.sav} for $k=11$.
 It must be noted that the number of
sites in the obtained equidistant lattice ($N\approx2\times10^4$)
is much larger than that in the generating Cantor set ($2^{11}$).
The inset is presented to illustrate the self-similarity of the
set more transparently.

\begin{figure}[!]
\begin{center}
\mbox{\epsfig{figure=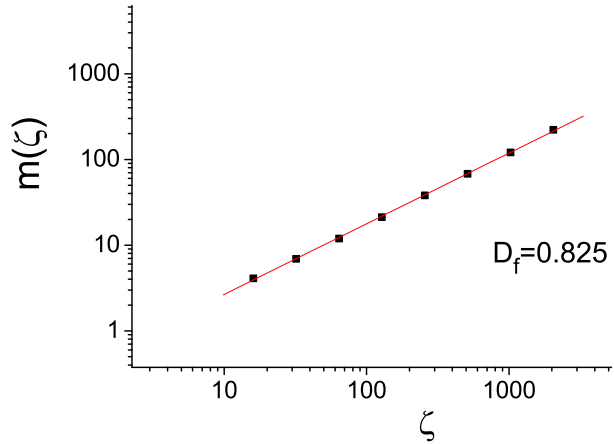 ,width=9cm,angle=0}}
\caption{\it $m(\zeta)$ vs $\zeta$, in the constructed set of
oscillators. The slope $D_f$ is equal to $5/6$ within an error of
less than 1\%.} \label{Det.power0}
 \end{center}
 \end{figure}

The constructed set of oscillators ($\s$-field) possesses the
property (\ref{fracmassdim}), as can be seen  in the log-log plot
of $m(\zeta)$ versus $\zeta$ presented in fig.~\ref{Det.power0}.
The exponent $D_f$, i.e the fractal mass dimension, is equal to
$5/6$ within an error of less than 1\%.

There are a few comments to be added concerning the connection of
the constructed oscillator chain with critical phenomena. In fact
the equilibrium position of the oscillators can be identified with
the order parameter of an equivalent critical system, while the
displacements $\s_\nu$ are associated with the fluctuations of the
order parameter. Clearly, at the critical point the expected
fluctuation pattern possesses fractal characteristics in close
analogy to the measure (\ref{fracmassdim}) defined on the
considered oscillator chain.

\section{Fractal measure deformation in the $1-D$ case}

In the previous section we determined the values of the scalar
$\s$-field on the $N$ lattice sites, in order to present fractal
characteristics. That is, from the total of $N$ oscillators we
have displaced $2^{k}$ of them to the value $\pm1$ while keeping
the rest to zero, in such a way that eq.~(\ref{fracmassdim}) holds
and the system possesses a fractal mass dimension $D_f$. We are
interested in studying the evolution of this $\s$-configuration
according to the dynamics determined by the Lagrangian
(\ref{eq:leap-froglagr}), and especially we focus on the evolution
of $m(\zeta)$. Before considering an anharmonic, in general,
potential suitable to describe the aforementioned symmetry
breaking (for example of fourth order), it is interesting to
investigate the simple harmonic case where there is some analytic
information, in order to acquire a better apprehension. However,
even this simple model will reveal a rich and unexpected behavior.

\subsection{Second order potential}

We consider first the second order potential
\begin{equation}
V(\s)=\frac{\la}{4}(\s-1)^2-A\s, \label{2pot}
\end{equation}
where $\la$ and $A$ are the coupling parameters of our model. All
the quantities ($\s$, $\la$, $A$, as well as the space-time
variables) appearing above are taken dimensionless.
 The corresponding equation of
motion for the $i$-th oscillator derived form
eqs.~(\ref{eq:leap-froglagr}) and (\ref{2pot}) is:
\begin{equation}
\ddot{\s}_i=\frac{1}{\alpha^2}
\left(\s_{i+1}+\s_{i-1}-2\s_{i}\right)-
\left[\frac{\la}{2}\s_{i}-\frac{\la}{2}-A\right].
\label{2leap-froga}
\end{equation}
In order to solve it we use the leap-frog  time discretization
scheme, leading to:
\begin{equation}
\s^{n+2}_i=2\s^{n+1}_i-\s^{n}_i+\frac{dt^2}{\alpha^2}
\left(\s^{n+1}_{i+1}+\s^{n+1}_{i-1}-2\s^{n+1}_{i}\right)-
dt^2\left[\frac{\la}{2}\s^{n+1}_{i}-\frac{\la}{2}-A\right],
\label{2leap-frogb}
\end{equation}
where $\alpha$ is the lattice spacing and $dt$ is the time step.
The upper indices indicate the time instants  and the lower
indices the lattice sites. As usual we perform an initial fourth
order Runge-Kutta step to make our algorithm self-starting, and we
impose periodic boundary conditions. The numerical integration
results are not sensitive to the $\alpha$ and $dt$ choice,
provided that $dt\leq\alpha/2$.

\begin{figure}[!]
\begin{center}
\mbox{\epsfig{figure=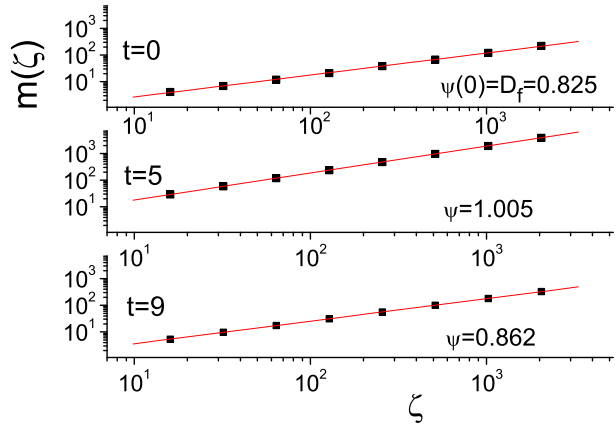,width=9cm,angle=0}}
\caption{\it $m(\zeta)$ vs $\zeta$ for three successive times,
$t=0$, $t=5$  and $t=9$, for $\la=1$ and $A=1$.}
\label{3sucslopes}
 \end{center}
 \end{figure}

Let us first assume zero initial kinetic energy, i.e
$\dot{\s}_i=0$ for every lattice site, which physically is a
strong requirement of equilibrium. We evolve the constructed
fractal configuration, obtained using $k=11$, according to
(\ref{2leap-frogb}) for various potential parameters. In
fig.~\ref{3sucslopes} we depict $m(\zeta)$ versus $\zeta$ for
three successive times, $t=0$, $t=5$ and $t=9$, for $\la=1$ and
$A=1$.
Initially the slope $\psi$ is the fractal mass dimension $D_f$.
Here we have chosen $D_f=5/6$. As we can see, the initial fractal
geometry is completely lost at $t=5$. However, at $t=9$ it is
almost re-established.

\begin{figure}[!]
\begin{center}
\mbox{\epsfig{figure=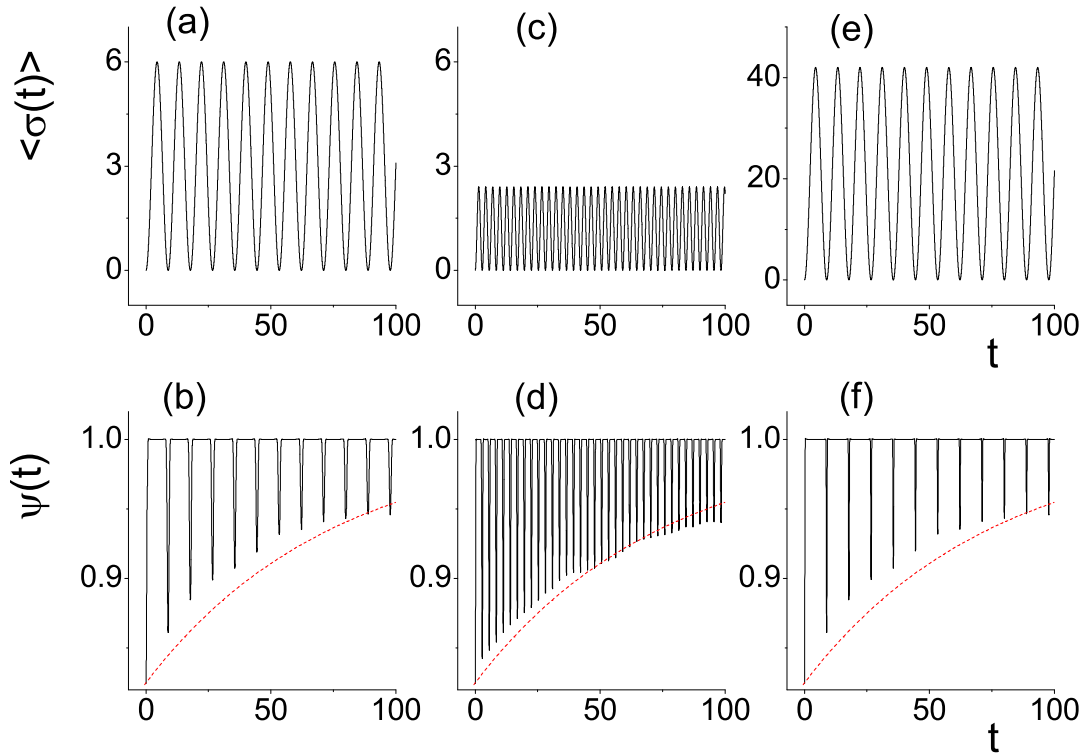,width=16cm,angle=0}}
\caption{\it $\langle\s(t)\rangle$ and $\psi(t)$ evolution for
$\la=1$ and $A=1$ ((a) and (b) plots), for $\la=10$ and $A=1$ ((c)
and (d) plots) and for $\la=1$ and $A=10$ ((e) and (f) plots). The
dashed line in the lower graphs marks the
$F(t)=1-(1-\psi(0))e^{-qt}$ curve, with $q=0.013$.}
\label{2nd.psit}
 \end{center}
 \end{figure}

Let us explore this remarkable result further. In the upper graphs
of fig.~\ref{2nd.psit} we present the evolution of the mean field
value $\langle\s(t)\rangle$ for three $\la$ and $A$ cases. In the
 lower ones we show the corresponding  evolution of the slope
$\psi(t)$ of $m(\zeta)$ versus $\zeta$  (each $\psi(t)$ value
obtained through a linear fit).
First of all, the spatial mean field value $\langle\s(t)\rangle$
oscillates around the potential minimum
$\s_{min}=1+\frac{2A}{\la}$ with constant amplitude  as expected.
Secondly, one oscillator moving in the potential (\ref{2pot}) has
the period $T=\frac{2\pi\sqrt{2}}{\sqrt{\la}}$, and this holds
equally well for $\langle\s(t)\rangle$ too, due to the
synchronization of the dominating zero background \cite{synchron}.

We observe that the exponent $\psi(t)$ rapidly reaches the
embedding dimension value 1, but it reacquires a value close to
the initial one periodically. It is easy to see that this (partial
but clear) re-establishment of the initial fractal geometry
happens at times where $\langle\s(t)\rangle$ returns to its
starting point, which is the lower turning point of the
oscillations. In the case we are looking at, this starting point
corresponds to $\langle\s(0)\rangle\approx0$ (since only a small
fraction of sites is displaced to $\pm1$ while the others form a
zero background) and to $\dot{\s}_i(0)=0$ for every oscillator.
Therefore, the explanation for this behavior is induced easily.
Indeed, initially only the discrete set of oscillators displaced
to $\pm1$ contributes to the integral (\ref{fracmassdim}), while
the zero background adds with zero effect. As the system of
coupled oscillators evolves in the potential (\ref{2pot}), this
zero background is excited and its non-zero but trivial
contribution to (\ref{fracmassdim}) sufficiently overcomes that of
the initial $\pm1$'s and consequently deforms completely the
fractal geometry. However, we expect a simultaneous return of this
background to zero (synchronization) since the energy transfer
between the different oscillators takes place through the spatial
derivative (which is small since the displacement to $\pm1$ is not
large compared to the potential minimum) and therefore only the
zeros close to the initial $\pm1$'s will return to a different
value. Moreover, this behavior is amplified by the initial zero
kinetic energy for all oscillators, which strengthens homogeneity.
As a result, at times where $\langle\s(t)\rangle$ and the zero
background return to the lower turning point, the system
re-exhibits a power law behavior in $m(\zeta)$ with exponent close
to the initial one, i.e to the fractal mass dimension $D_f=5/6$.
Each re-appearance of the initial fractal geometry will survive as
long as the system stays close to its lower turning point,
therefore the corresponding interval will be larger for smoother
potentials at their minimum. This effect can be weakly seen
comparing the lower left and right plots of fig.~\ref{2nd.psit}.
Lastly, due to the non-trivial excitation of the zeros in the
neighborhood of the initial $\pm1$'s, which number increases
monotonically as time passes, every partial re-establishment of
the initial fractal geometry will possess slightly larger exponent
than the previous one. This behavior is observed in
fig.~\ref{2nd.psit}, where the $\psi(t)$ value at the minima
increases successively. Therefore, we expect that the dynamics
will totally deform the original fractality in the end.

\begin{figure}[!]
\begin{center}
\mbox{\epsfig{figure=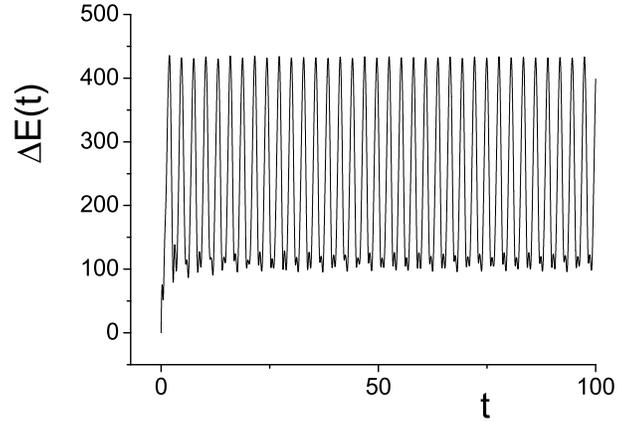,width=9cm,angle=0}} \caption{\it
$\triangle E(t)$ for $\la=10$ and $A=1$. The minima coincide with
those of $\langle\s(t)\rangle$ and $\psi(t)$.} \label{2nd.DEt}
 \end{center}
 \end{figure}

A supporting argument for the cogitation analyzed above is the
calculation of $\triangle
E(t)=\sqrt{\sum_i^N\left[E_i(t)-E_i(0)\right]^2}$, which provides
a measure for the divergence of the oscillators' total energies
from their initial values. In fig.~\ref{2nd.DEt}  we plot
$\triangle E(t)$ for $\la=10$ and $A=1$ case.
As we observe, it presents minima for the same times as
$\langle\s(t)\rangle$ and $\psi(t)$, therefore the partial
re-appearance of the fractal geometry happens when the oscillators
acquire energies close to their initial ones. However, $\triangle
E(t)$ cannot describe the mixing of energy between the different
oscillators.

There are two time scales in the reappearance phenomenon. The
first, naming $\tau_1$, is the period of the partial
re-establishment of the initial fractal mass dimension. It
coincides with the oscillations period as we have already
mentioned. The second, $\tau_2$, is the time scale which
determines the complete deformation of the initial fractal
geometry. We quantify an estimation of $\tau_2$ by assuming that
the ratio of two successive variations of $\psi(t)$ at the
corresponding minima is constant. If $\psi_l$ denote these
successive minima and $t_l=l\tau_1$ the corresponding successive
times, this natural assumption reads:
\begin{equation}
\frac{\psi_{l+2}-\psi_{l+1}}{\psi_{l+1}-\psi_{l}}=C.
\label{qfracfit}
\end{equation}
This finite difference equation has the solution:
\begin{equation}
\psi_{l}=1-(1-\psi(0))\,e^{-qt_l}, \label{qfit}
\end{equation}
if $C=e^{-q\tau_1}$, where we have added the necessary terms in
order to get correct values for $l=0$ ($\psi(0)$) and for
$l\rightarrow\infty$ ($\psi(t_l)\rightarrow1$). Therefore, we
determine $\tau_2$ in terms of the exponent $q$ as
$\tau_2\approx5/q$, since the characteristic time is $1/q$ and for
$t=5/q$ the deviation from the embedding dimension is considered
as completely lost, falling below the threshold value of $10^{-2}$
(in close analogy with capacitor discharge in electronics). The
approximation (\ref{qfracfit}) reproduces very well the exact
$\psi_l$ behavior, as can be observed in fig.~\ref{2nd.psit} where
we display, with the dashed line, the analytical estimation
(\ref{qfit}) for $q=0.013$. Indeed, $q$ offers a measure of the
deformation of the initial fractal geometry, and in the following
we investigate numerically its dependence on the various model
parameters.

Firstly, $q$, i.e $\tau_2$, is completely independent from $\la$
and $A$, contrary to $\tau_1$ which (coinciding with the
oscillation period) depends on $\la$. The constant value of $q$
for different $A$ and $\la$, is a result of the increase of the
oscillation frequency combined with a compensating decrease of the
difference $\psi_{l+1}-\psi_{l}$. The corresponding $q-\la$ and
$q-A$ plots are trivial horizontal lines.

\begin{figure}[!]
\begin{center}
\mbox{\epsfig{figure=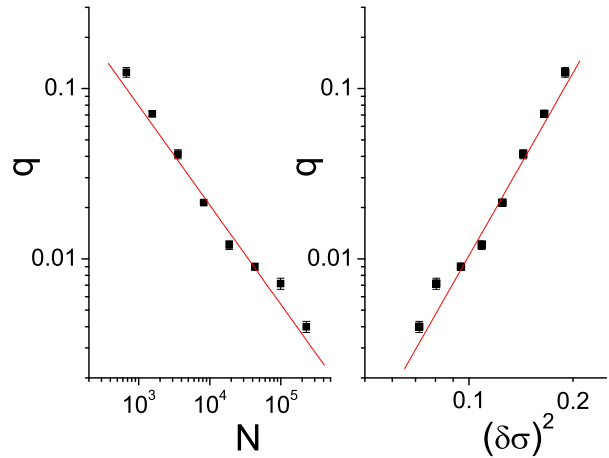,width=9cm,angle=0}}
\caption{\it On the left we show the exponent $q$ defined in
eq.~(\ref{qfit}) versus total lattice site number $N$, for $\la=1$
and $A=1$. On the right we show $q$ versus the initial variation
of the field values $(\delta\s)^2$, calculated for the different
$N$ values of the left plot. The solid lines mark exponential
fits.} \label{2nd.qNds2}
 \end{center}
 \end{figure}

On the left plot of fig.~\ref{2nd.qNds2} we depict the dependence
of $q$ on the total number of lattice sites $N$, for $\la=1$ and
$A=1$.
 Note that increasing $k$ in the construction of the Cantor $2^k$-point set, we result
to a much more rapidly growing equidistant lattice (the points in
fig.~\ref{2nd.qNds2} correspond to a successive increase of $k$
from 7 to 13). It is clear that $q$ decreases exponentially with
$N$, therefore the deformation becomes weaker. The explanation is
straightforward since by increasing the number of the Cantor
points (which will form the $\pm1$'s in the equidistant lattice)
we need much more sites with zero value. In other words, the
measure of the $\pm1$'s relatively to the zero background
decreases with $N$. Due to the influence of this background to the
re-appearance phenomenon (larger times needed for the mixing of
the oscillators through the spatial derivative) the deformation
becomes weaker with $N$, i.e $q$ decreases. Finally, it is obvious
that in an infinite system the initial fractal geometry will be
periodically deformed and re-established for infinite time
($q\rightarrow0$, i.e $\tau_2\rightarrow\infty$ for
$N\rightarrow\infty$), since in this case the initial measure of
the $\pm1$'s is zero and therefore infinite time is needed for the
mixing and de-synchronization of the zero background.

On the right graph of fig.~\ref{2nd.qNds2} we show the dependence
of $q$ on the initial variation of $\s$-field, calculated by
$(\delta\s)^2=\langle \s^2\rangle-\langle
\s\rangle^2=\sum_i\s_i^2/N-(\sum_i\s_i/N)^2$ at $t=0$, for the
different $N$ values used in the left plot. The initial variation
of the $\s$-field, reflecting the domination of the homogenous
zero background, affects the deformation exponent $q$. Larger
initial $(\delta\s)^2$ values correspond to zero background with
smaller measure relatively to the $\pm1$'s, and therefore to
weaker re-appearances, i.e to larger $q$'s. On the other hand, for
$N\rightarrow\infty$ $(\delta\s)^2\rightarrow0$, the measure of
the $\pm1$'s becomes zero, and the re-appearance phenomenon holds
for ever ($q\rightarrow0$, i.e $\tau_2\rightarrow\infty$).

\begin{figure}[!]
\begin{center}
\mbox{\epsfig{figure=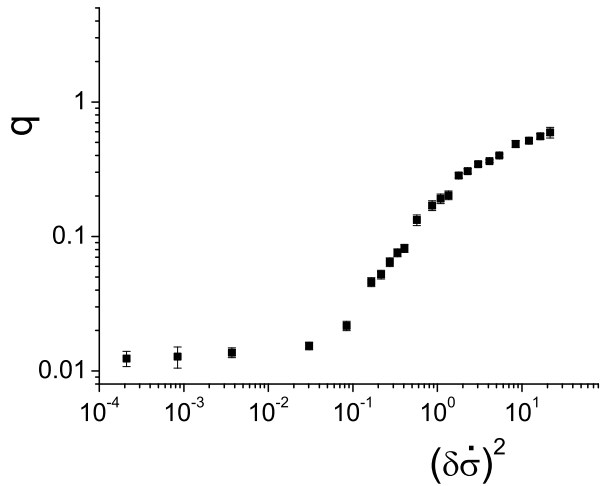,width=9cm,angle=0}}
\caption{\it Exponent $q$ versus initial variation of time
derivatives of the field values $(\delta\dot{\s})^2$, for
$N=18819$ ($2^{11}$ Cantor points), for $\la=1$ and $A=1$.}
\label{2nd.q.sdot}
 \end{center}
 \end{figure}

Another possibility could be to change the initial $(\delta\s)^2$
by displacing randomly all the oscillators from their constructed
values, while keeping $N$ constant. However, we avoid doing so
since this procedure alters the initial fractal mass dimension.
Instead, we may perturb randomly the initial time derivatives and
investigate the effect of the variation of the initial kinetic
energies on $q$. In fig.~\ref{2nd.q.sdot} we present the
dependence of $q$ on the variation of $\dot{\s}_i$ at $t=0$, given
by $(\delta\dot{\s})^2=\langle \dot{\s}^2\rangle-\langle
\dot{\s}\rangle^2=\sum_i\dot{\s}_i^2/N-(\sum_i\dot{\s}_i/N)^2$,
for $N=18819$ ($2^{11}$ Cantor points), in the $\la=1$, $A=1$
case. Indeed we observe a significant increase of $q$ for larger
$(\delta\dot{\s})^2$ as expected, due to the de-synchronization of
the zero background, i.e the initially zero oscillators are
excited and mixed due to their different kinetic energies too,
apart from their coupling to the $\pm1$'s. Mind that the turning
point is not zero any more and
 $\langle\s(t)\rangle$ moves to negative values, too.
 Finally, note that additionally to
this $\dot{\s}(0)$ perturbation, there is always a constant
initial $(\delta\s)^2$ present, resulting from the $\pm1$'s
($(\delta\s)^2\approx0.1$ in this specific case), which cannot be
removed. Therefore, the stabilization of $q$ for sufficiently
small $(\delta\dot{\s})^2$ is due to the overcoming effect of
$(\delta\s)^2$ comparing to that of $(\delta\dot{\s})^2$.

Finally, a quantitative measure concerning the aforementioned
dynamics is the Lyapunov exponent. Following \cite{Pettini} we can
calculate it analytically in this simple second order potential
case. The constant curvature of the external potential leads to
Lyapunov exponent exactly equal to zero, which according to our
analysis can be related to $\lim_{N\rightarrow\infty}q$. However,
a proof of this statement goes beyond the scope of the present
work.

\subsection{Fourth order potential}

After analyzing the simple second order potential case, which
revealed an interesting behavior though, we extend our
investigation to the fourth order model which will give rise to
non-linear equations of motion. The potential has the form
\begin{equation}
V(\s)=\frac{\la}{4}(\s^2-1)^2-A\s. \label{4pot}
\end{equation}
 Inspired by
the $\s$-model we assume that the $Z_2$ symmetry
($\s\rightarrow-\s$) is broken only through a linear term, setting
the coefficient of the cubic term in the potential to zero.
 The  equations of
motion derived form  eqs.~(\ref{eq:leap-froglagr}) and
(\ref{4pot}) is:
\begin{equation}
\ddot{\s}_i=\frac{1}{\alpha^2}
\left(\s_{i+1}+\s_{i-1}-2\s_{i}\right)-
\left[\lambda\s_{i}^3-\lambda\s_{i}-A\right],~~~~~~~~~i=1,\ldots,N
\label{4leap-frog}
\end{equation}
which are solved using the leap-frog  time discretization
algorithm given in the previous subsection.

\begin{figure}[h]
\begin{center}
\mbox{\epsfig{figure=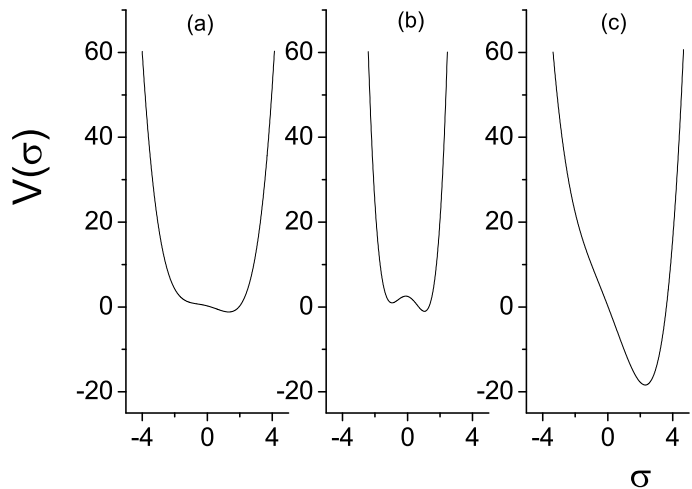,width=10.1cm,angle=0}}
\caption{\it Potential (\ref{4pot}) for $\la=1$ and $A=1$ (plot
(a)), for $\la=10$ and $A=1$ (plot (b)) and for $\la=1$ and $A=10$
(plot (c)).} \label{4rth.potential}
\end{center}
\end{figure}

\begin{figure}[!]
\begin{center}
\mbox{\epsfig{figure=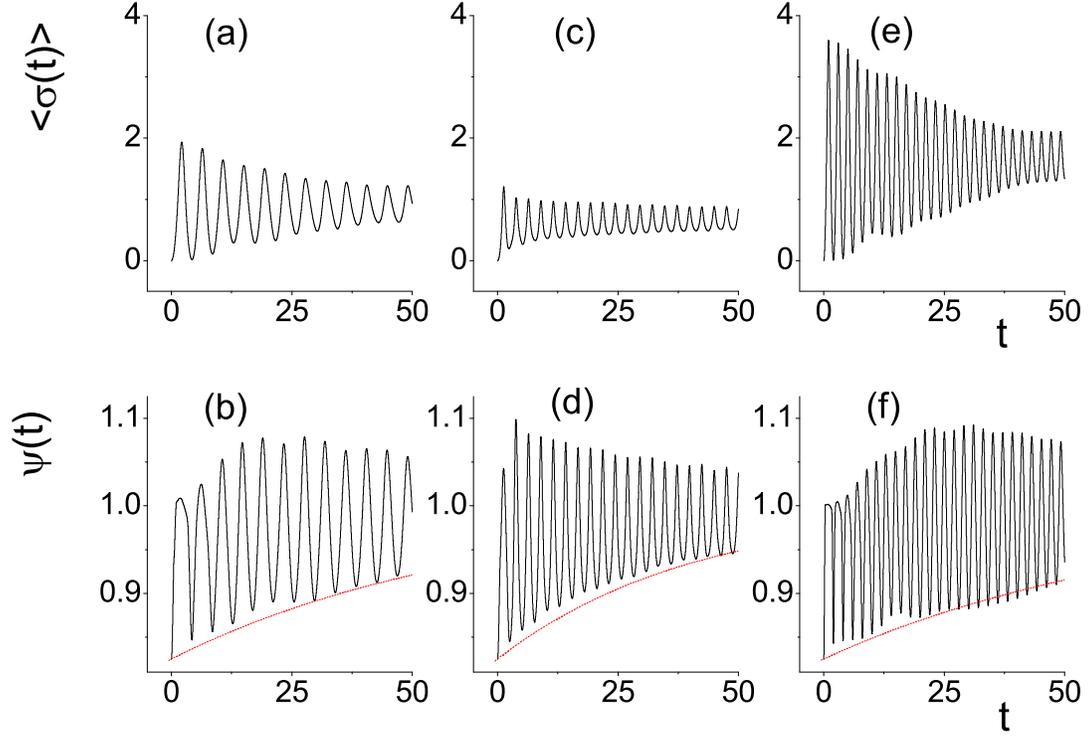,width=16cm,angle=0}}
\caption{\it $\langle\s(t)\rangle$ and $\psi(t)$ evolution for
$\la=1$ and $A=1$ ((a) and (b) plots), for $\la=10$ and $A=1$ ((c)
and (d) plots) and for $\la=1$ and $A=10$ ((e) and (f) plots), in
the fourth order potential case. The dashed line in the lower
graphs marks the $F(t)=1-(1-\psi(0))e^{-qt}$ curve, with
$q=0.016$, $q=0.024$ and $q=0.014$, respectively.}
\label{4rth.psit}
\end{center}
\end{figure}

We evolve the constructed fractal configuration according to
(\ref{4leap-frog}), assuming zero initial kinetic energy, for
various potential parameters. In fig.~\ref{4rth.potential} we draw
the potential for three $\la$ and $A$ cases and in
fig.~\ref{4rth.psit} the corresponding evolution of
$\langle\s(t)\rangle$ and $\psi(t)$.
The spatial mean field value $\langle\s(t)\rangle$ oscillates
around the potential minimum, which now is one of the three roots
of $V'(\s)=\la\s^3-\la\s-A=0$. Depending on $\la$ and $A$ we can
have two minima and a maximum (middle plot of
fig.~\ref{4rth.potential}), one minimum and one saddle point, or
just one minimum (left and right plots of
fig.~\ref{4rth.potential}). Contrary to the second order case of
fig.~\ref{2nd.psit}, the oscillation amplitude decreases with time
due to the anharmonic dynamics. However, the amplitude attenuation
weakens with increasing lattice size $N$ and for an infinite
system it remains constant.

\begin{figure}[!]
\begin{center}
\mbox{\epsfig{figure=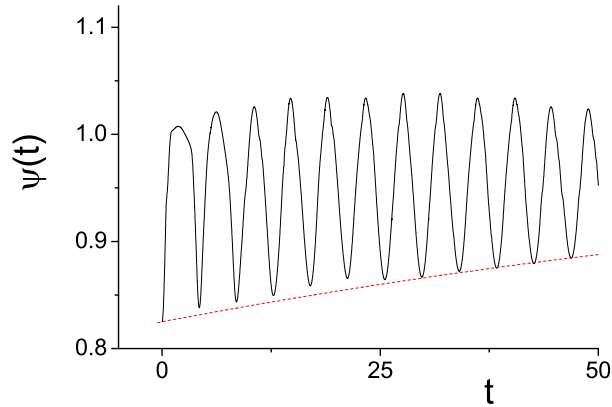,width=9cm,angle=0}}
\caption{\it $\psi(t)$ evolution for $\la=1$ and $A=1$ in the
fourth order potential case, for a Cantor-like lattice of size
$N=228208$, obtained using $k=14$. The dashed line marks the
$F(t)=1-(1-\psi(0))e^{-qt}$ curve, with $q=0.0089$. The effective
exceeding of 1 is decreased due to the increased lattice size.}
\label{4rth.N16384}
\end{center}
\end{figure}

\begin{figure}[!]
\begin{center}
\mbox{\epsfig{figure=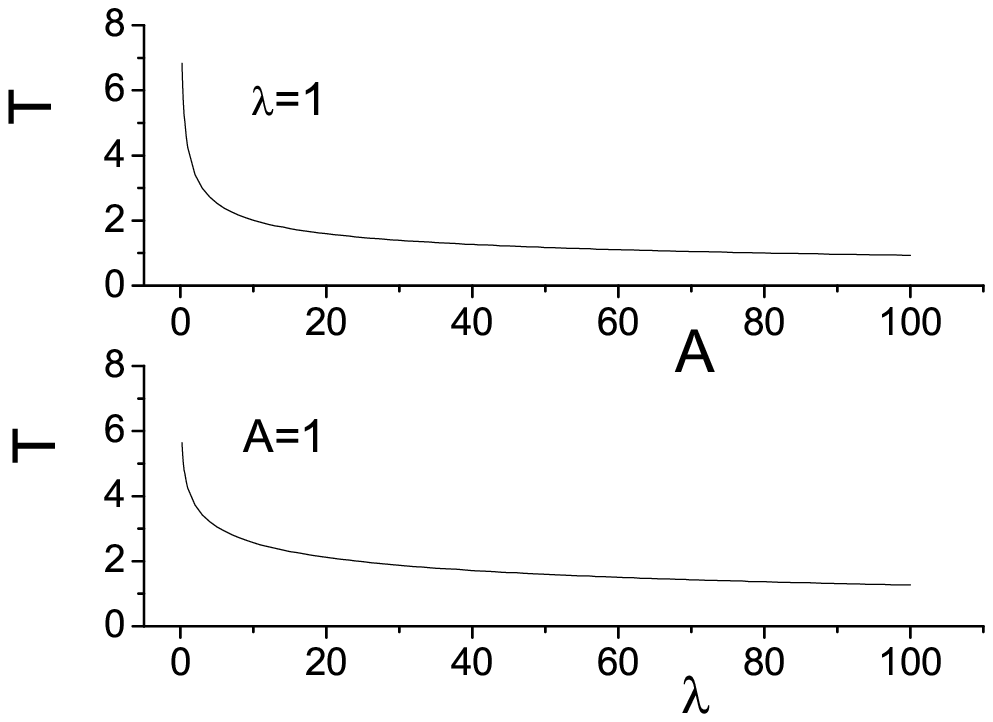,width=9cm,angle=0}}
\caption{\it Dependence of oscillations period $T$, which
coincides with time scale $\tau_1$, on $A$ and $\la$ in the fourth
order potential case.} \label{4rth.period}
 \end{center}
 \end{figure}

As we observe in fig.~\ref{4rth.psit}, the periodical partial
re-appearance of the initial fractal mass dimension at times when
$\langle\s(t)\rangle$ has a minimum, holds similarly to the
harmonic case. Firstly, in this anharmonic case, as $\psi(t) \to
1$ the power-law form of $m(\zeta)$ is slightly distorted at large
$\zeta$ values due to the finite system size. Consequently, the
power-law fit leads occasionally to effective $\psi(t)$ slightly
greater than 1. To correct this behavior, one could restrict the
fit to smaller $\zeta$ values (worsening statistics) or go to
significantly larger lattices (huge computational times). Indeed,
in fig.~\ref{4rth.N16384} we present $\psi(t)$ evolution for a
Cantor-like lattice of size $N=228208$, obtained using $k=14$,
where the tendency of $\psi(t)$ to approach the limiting value
($\psi(t) \to 1$) becomes obvious. However, since this effect does
not influence the re-appearance phenomenon and its
characteristics, we will ignore it in the following analysis.
 Secondly, the time
scale $\tau_1$, which determines the period of the re-appearance,
coincides with the oscillation period $T$. It cannot be calculated
analytically and in fig.~\ref{4rth.period} we depict its
dependence on $\la$ and $A$ found numerically.
Thirdly, the computation of $\triangle E(t)$ leads to similar to
the second order case results, that is it possesses minima
simultaneously with $\langle\s(t)\rangle$ and $\psi(t)$.

\begin{figure}[!]
\begin{center}
\mbox{\epsfig{figure=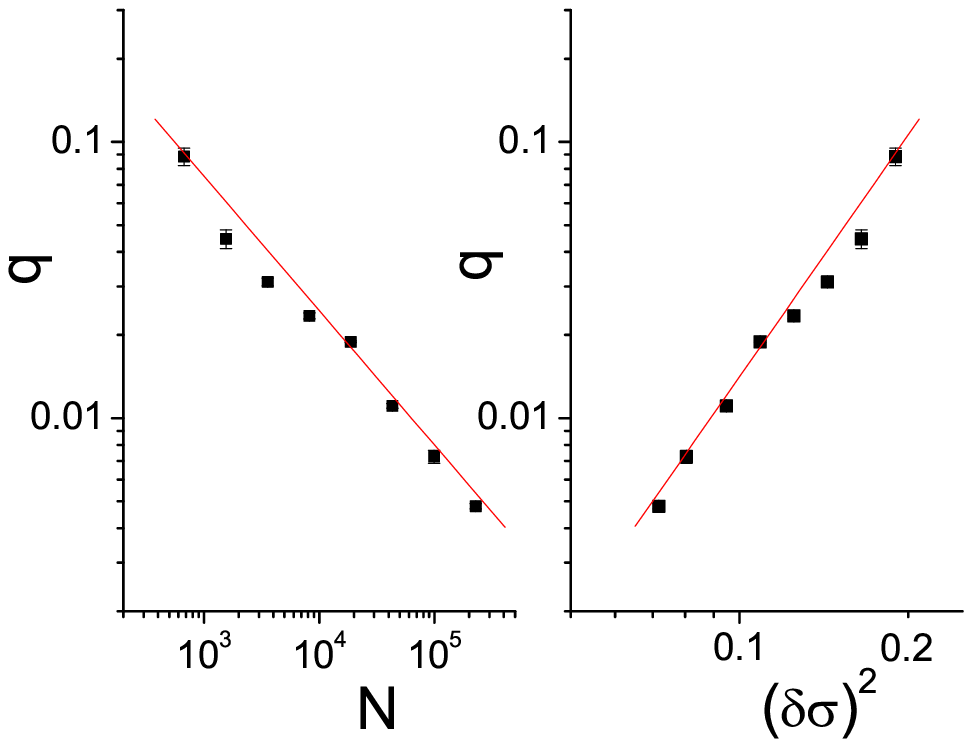,width=9cm,angle=0}}
\caption{\it On the left, the exponent $q$ and its errors versus
total lattice site number $N$, for $\la=1$ and $A=1$ in the fourth
order potential case. On the right, $q$ versus the initial
variation of the field values $(\delta\s)^2$, calculated for the
different $N$ values of the left plot. The solid lines mark
exponential fits.} \label{4rth.qNds2}
 \end{center}
 \end{figure}

Continuing we study the dependence of the exponent $q$ of
eq.~(\ref{qfit}), which quantifies the gradual permanent
deformation of the initial fractal geometry, on the various
parameters. Note however that in general in this fourth order
potential case the upper envelope of the $\psi(t)$ graph is more
complex. In fig.~\ref{4rth.qNds2} we present $q$ versus lattice
site number $N$ (left plot), and versus the corresponding initial
variation $(\delta\s)^2$ (right plot).
Its behavior is similar to the second order case of
fig.~\ref{2nd.qNds2} and the interpretation is the same. However,
the corresponding $q$ values seem to be slightly increased, that
is the anharmonic dynamics deforms the initial fractal geometry
earlier. The explanation of this behavior is the decreasing
oscillations amplitude of this case (see fig.~\ref{4rth.psit}).
Indeed, the system lower turning point moves gradually to larger
values, i.e $\langle\s(t)\rangle$ does not return to zero and the
initially zero background remains excited, thus spoiling the
fractal mass dimension. The permanent oscillators displacement
from zero is an additional mechanism of the fractality deformation
in a finite system, apart from the mixing and de-synchronization
caused by the partial derivative. Its effect weakens with
increasing $N$, since the amplitude attenuation weakens too, as we
have already mentioned.

\begin{figure}[!]
\begin{center}
\mbox{\epsfig{figure=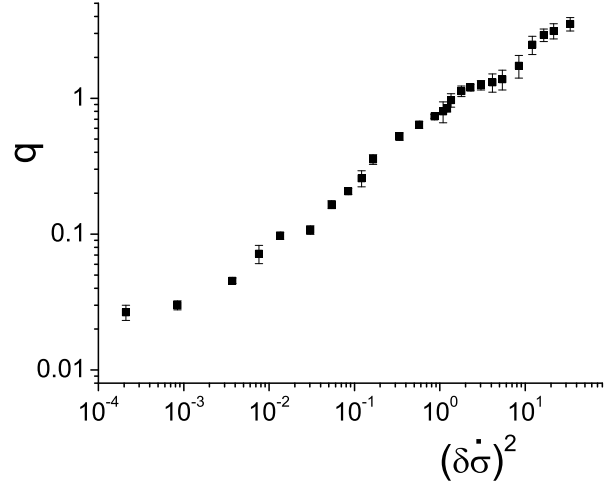,width=9cm,angle=0}}
\caption{\it Exponent $q$ versus initial variation of time
derivatives of the field values $(\delta\dot{\s})^2$, for
$N=18819$ ($2^{11}$ Cantor points), for $\la=1$ and $A=1$ in the
fourth order potential case.} \label{4rth.q.sdot}
 \end{center}
 \end{figure}

The amplified deformation rate can be deduced also from
fig.~\ref{4rth.q.sdot}, where we depict the dependence of $q$ on
the initial variation of $\dot{\s}_i$. It resembles the
corresponding fig.~\ref{2nd.q.sdot} of the harmonic case but now
$q$ is significantly larger, especially for large
$(\delta\dot{\s})^2$.
Therefore, the increased initial kinetic energy interferes
intensely with the complex fourth order dynamics, leading to a
deformation of the initial fractality at significantly smaller
times.

\begin{figure}[!]
\begin{center}
\mbox{\epsfig{figure=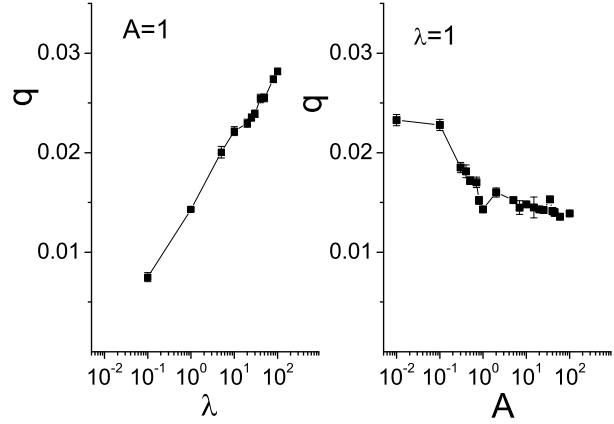,width=9cm,angle=0}}
\caption{\it $q$ dependence on $\la$ (left plot) and $A$ (right
plot) in the fourth order potential case, for $N=18819$ ($2^{11}$
Cantor points).} \label{4rth.qala}
\end{center}
\end{figure}

The main difference between fourth and second order cases, is the
effect of $\la$ and $A$ on $q$. Contrary to the previous harmonic
potential, where $q-\la$ and $q-A$ plots are trivial horizontal
lines, in fig.~\ref{4rth.qala} we show these graphs for the case
in hand.
We elicit that $q$ increases almost algebraically with $\la$ while
it decreases with $A$ in a more complex way. Although variation of
$\la$ seems to be slightly more important than that of $A$, both
have less influence on $q$ than $N$ and $\dot{\s}_i$.

A possible explanation of this dependence of $q$ on $\la$ and $A$,
could be the corresponding Lyapunov exponent. For the fourth order potential
this exponent cannot be calculated analytically. We estimate it semi-analytically
following \cite{Pettini}, and we find that for finite $N$ it is
not zero anymore, but it takes a small non-zero value depending on
the potential parameters, especially on its curvature, i.e on
$\la$. However, even in this anharmonic case, the Lyapunov
exponent seems to tend to zero for larger $N$, therefore an
infinite system will posses the re-appearance phenomenon for
infinite time ($q\rightarrow0$, i.e $\tau_2\rightarrow\infty$ for
$N\rightarrow\infty$), consistently with fig.~\ref{4rth.qNds2}.

\section{Two and higher dimensional evolution}

It is necessary to investigate the validity of the allegation
described above in higher dimensional systems, where critical
behavior can naturally arise. Keeping as a central observable of
interest for the critical system the fractal geometry of the
clusters formed at the critical point, it is possible to model, in
a simplified manner, the critical system with an ordinary
geometrical set possessing the appropriate fractal mass dimension.
In fact we can construct a set with dimension $D_f$ embedded in a
$D$-dimensional space, by taking the Cartesian product of $1-D$
sets, generated by the procedure described in section II, each one
having fractal mass dimension $D_f/D$ \cite{Falconer}. For
simplicity we consider here the $2-D$ case, leading  to a
$N_1\times N_2$ lattice where the field values are the products of
the corresponding one-dimensional ones, thus resulting to
$2^{k_1}\times 2^{k_2}$ $\pm1$'s. As a concrete example, following
the steps of section II we produce a $1551\times1551$ lattice
arising from the Cartesian product of two $2^8$ Cantor sets, each
one possessing fractal mass dimension $5/6$. The  set of lattice
sites with non-vanishing field values is a finite realization of a
fractal set with dimension $5/3$ embedded in a two dimensional
space.

\begin{figure}[!]
\begin{center}
\mbox{\epsfig{figure=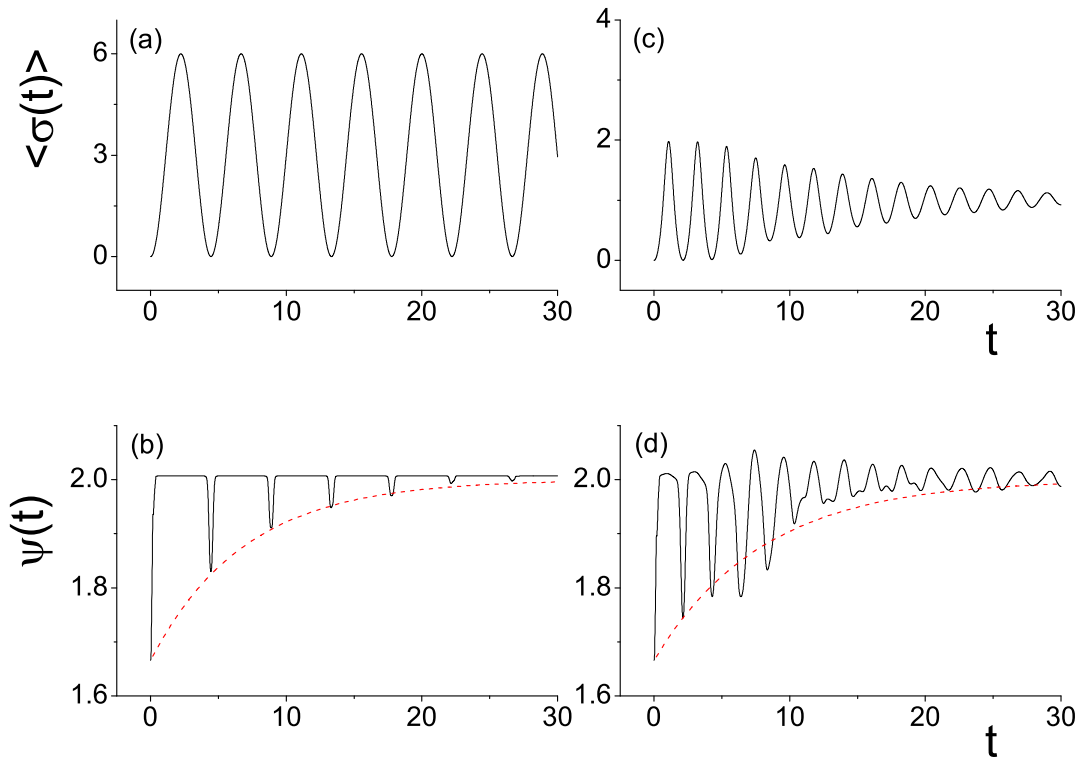,width=16cm,angle=0}}
\caption{\it $\langle\s(t)\rangle$ and $\psi(t)$ evolution for
$\la=1$ and $A=1$, for the second order potential ((a) and (b)
plots), and for the fourth order one ((c) and (d) plots). The
dashed line in the lower graphs marks the
$F(t)=2-(2-\psi(0))e^{-qt}$ curve, with $q=0.14$ and $q=0.13$
respectively.} \label{2D.fields.psit}
\end{center}
\end{figure}

It is straightforward to generalize the equations of motion
(\ref{2leap-froga}), (\ref{4leap-frog}), for the second and fourth
order potential cases respectively. Initial equilibrium
corresponds to field configuration with zero kinetic energy,
similarly to the  $1-D$ analysis. The evolution of the system is
depicted in fig.~\ref{2D.fields.psit}
for $\la=1$ and $A=1$. We show the mean displacement
$\langle\s(t)\rangle$ (averaged over the lattice), as well as the
running mass dimension $\psi(t)$. We observe the same phenomenon
of the partial re-establishment of the initial fractal geometry
every time $\langle\s(t)\rangle$ approaches its lower turning
point. The time scale $\tau_1$ of the re-appearance period
coincides with that of the oscillations, and the envelope of the
minima of $\psi(t)$ has an exponential form with exponent $q$
similarly to the $1-D$ case, suggesting that the analysis of the
previous section is also valid in this case. It is clearly seen in
fig.~\ref{2D.fields.psit} that $\psi(0)$ equals $5/3$, and
$\psi(t)$ reaches successively the embedding dimension value 2 as
expected. However the exponent $q$ is almost one order of
magnitude larger, leading to the conclusion that the higher
dimensional dynamics deforms the initial fractal geometry earlier.

The same procedure can be easily extended to three dimensions.
However since the lattice  site number increases rapidly with
increasing dimension we have to use a very coarse-grained
approximation of the initial Cantor set in order to acquire
plausible computational evolution times.

\section{Summary and conclusions}

In the present work we have investigated the evolution of a
fractal set resembling the order parameter clusters formed at the
critical point of a macroscopic system. Our analysis is based on a
simplified description of the critical system, restricted to the
reproduction of the correct fractal mass dimension. We assumed
initial equilibrium and we explored the variation of this
appropriately defined fractal mass dimension with time. We have
found that the initial fractal geometry is being deformed and
partially re-established periodically, at times when the mean
field value returns to its lower turning point. The origin of this
effect is made more transparent in a harmonic $1-D$ model. For a
complete study we investigated the influence of anharmonic
interactions as well as initial deviations form equilibrium, on
the time scales determining the re-appearance phenomenon. We
derive an analytical expression describing, to a sufficient
accuracy, the value of the running fractal mass dimension
$\psi(t)$ at the re-appearance times, and we show that the
re-appearance frequency coincides with that of the oscillations.
The total duration of the re-establishment process is inversely
proportional to a characteristic exponent $q$, which depends on
various parameters of the model. In particular $q$ is a decreasing
function of the total lattice size $N$, the initial field
variation $(\delta\s)^2$ and the initial time derivative variation
$(\delta\dot{\s})^2$. Therefore in an infinite system the initial
fractal mass dimension re-appears for ever. The only qualitative
difference of the harmonic and anharmonic analysis is the $q$
dependence on the potential parameters $\la$ and $A$ in the fourth
order case.

The same treatment can be followed in higher dimensional
($D\geq2$) systems, too. In these more relevant, for the
simulation of real critical systems, cases, we observe a similar
behavior which can be explained in an analogous manner. The only
quantitative difference is that $q$ increases significantly with
$D$. Therefore, the partial re-appearance of the initial fractal
geometry seems to be a robust property of the evolution of
critical systems, rendering the corresponding fractal dimension a
significant observable which can be determined even in the
symmetry broken phase. This in turn allows for the calculation of
critical exponents and the determination of the universality class
of the occurring transition. Our analysis is of interest for the
study of the fireball evolution in a heavy-ion collision
experiment, when the system at some intermediate stage passes
through the QCD critical point, and the main question is whether
imprints of the transient critical state can sustain for
sufficiently large times in order to be observed at the detectors
\cite{manosqcd}. Similarly, it could be applied to the primordial
fractal fluctuations of the inflationary field in order to
investigate their evolution to the present large scale
inhomogeneities and, with appropriate modifications, to the
evolution of astrophysical fractals, such as star and galaxy
clusters, to determine their deformation scales.\\

\paragraph*{{\bf{Acknowledgements:}}} We thank V. Constantoudis and
N. Tetradis for useful discussions. One of us (E.N.S) wishes to
thank the Greek State Scholarship's Foundation (IKY) for financial
support. The authors acknowledge partial financial support through
the research programs ``Pythagoras'' I and II of the EPEAEK II
(European Union and the Greek Ministry of Education) and
``Kapodistrias'' of the University of Athens.

\end{document}